\documentclass[pre]{revtex4}
\usepackage{amsmath}
\usepackage{graphicx}

\begin{document}
\title{The Cosmic Causal Mass}

\author{Simen Braeck}
\email{Simen.Brack@hioa.no}
\affiliation{Oslo and Akershus University College of Applied Sciences, Faculty of Technology, Art and Sciences,
\mbox{PB 4 St. Olavs. Pl.,} NO-0130 Oslo, Norway}

\author{\O yvind~G.~Gr\o n}
\email{oyvind.gron@hioa.no}
\affiliation{Oslo and Akershus University College of Applied Sciences, Faculty of Technology, Art and Sciences,
\mbox{PB 4 St. Olavs. Pl.,} NO-0130 Oslo, Norway}

\author{Ivar Farup}
\email{ivar.farup@ntnu.no}
\affiliation{Department of Computer Science, NTNU-Norwegian University of Science and Technology, \mbox{2815 Gj\o vik, Norway}}

\begin{abstract}
In order to provide a better understanding of rotating universe models, and in particular the G\"{o}del universe, we discuss the relationship between cosmic rotation and perfect inertial dragging. In this connection, the concept of \emph{causal mass} is defined in a cosmological context, and discussed in relation to the cosmic inertial dragging effect. Then, we calculate the mass inside the particle horizon of the flat $\Lambda$CDM-model integrated along the past light cone. The calculation shows that the Schwarzschild radius of this mass is around three times the radius of the particle horizon. This~indicates that there is close to perfect inertial dragging in our universe. Hence, the calculation provides an explanation for the observation that the swinging plane of a Foucault pendulum follows the stars.
\end{abstract}

\maketitle

\section{Introduction \label{sect:sec1}}

Here, we shall investigate a consequence of the general theory of relativity in the spirit of Einstein when he presented the general theory of relativity in his great article that was published in May 1916~\cite{B1}, and whose centenary was last year~\cite{B2}.

The extension of the principle of relativity from rectilinear motion with constant velocity to accelerated (both rectilinear and rotational) motion was an important motivating factor for Einstein when he constructed the general theory of relativity. He was inspired by the point of view of E. Mach who wrote in 1872: ``It does not matter whether we think of the Earth rotating around its axis, or if we imagine a static Earth with the celestial bodies rotating around it."~\cite{B3} Mach further wrote: ``Newton’s experiment with the rotating vessel of water simply informs us that the relative rotation of the water with respect to the sides of the vessel produces no noticeable centrifugal forces, but that such forces are produced by its relative rotation with respect to the mass of the Earth and the other celestial bodies."~\cite{B4}.

The second paragraph of Einstein’s great 1916 article is titled: The Need for an Extension of the Postulate of Relativity. He starts by writing that the restriction of the postulate of relativity to uniform translational motion is an inherent epistemological defect. Then, he writes: ``\emph{The laws of physics must be of such a nature that they apply to systems of reference in any kind of motion}. Along this road, we arrive at an extension of the postulate of relativity".

This point of view was inspired by the conception that if a particle is alone in the universe, then one cannot decide whether it moves or not. Hence, all motion should be relative, meaning that an observer cannot decide whether he is at rest or in motion. An arbitrary observer may consider himself as at rest. Note, however, that when an observer has a translational acceleration or a rotation relative to the average cosmic mass, she will experience, what in Newton’s theory is called, fictitious forces; but that is considered to be a gravitational effect, namely an effect of inertial dragging, according to Einstein’s~theory.

This means that it should be a valid point of view for an observer on the Earth to consider himself as at rest and the universe as moving. This requires that the observer be able to explain all his experiments under the assumption that he is at~rest.

P Kerzberg~\cite{B5} has given a very interesting discussion of ``the Relativity of Rotation in the Early Foundations of General Relativity". In particular, he reviews and comments on an article published in 1917 titled On the Relativity of Rotation in Einstein’s Theory by W. De Sitter~\cite{B6}. Kerzberg writes ``De Sitter thus maintains that rotation is relative in Einstein’s theory, and even as relative as linear translation. Both rotation and translation are susceptible to being transformed away. Nonetheless a difference persists". He then cites De Sitter: ``If a linear translation is transformed away (\mbox{by a Lorentz} transformation), it is utterly gone; no trace of it remains. Not so in the case of rotation. The~transformation which does away with rotation, at the same time alters the equation of relative motion in a definite manner. This shows that rotation is not a purely kinematical fact, but an essential physical reality."

According to the general theory of relativity, there is a connection between the phenomenon of inertial dragging, also called the Lense--Thirring effect~\cite{B7}, and the possible validity of the principle of relativity for rotating motion in our universe. A nice review of the history of the Lense--Thirring effect has been given by Pfister~\cite{B8}. Past, ongoing and future attempts to measure it in the Earth’s gravitational field and other astronomical systems are reviewed in~\cite{B9,B10} and references therein. It~should be noted that, recently, attempts to empirically detect/constrain the effects of rotation of a distant shell on Solar System planets were undertaken with recent planetary ephemerides~\cite{B11,B12}.

A few years ago, C. H. Brans~\cite{B13} presented an interesting discussion of ``What Exactly is ``Mach’s Principle?"". He wrote that ``it is probably safe to say that Mach’s Principle seeks a physical model/theory which explains/predicts that inertial reference frames are those which see the fixed stars to have zero acceleration". This is valid in the context of the usual non-rotating Friedmann--Lemaître--Robertson--Walker (FLRW) universe models, but it is not valid for rotating universe models, such as the G\"{o}del model, which will be discussed in the present paper. What is generally valid is that local inertial reference frames are those in which free particles will not accelerate, and the swinging plane of a pendulum will not rotate. However, in a rotating universe model, an~observer at rest in an inertial reference frame will observe  that ‘the distribution of the fixed stars’~rotates.

Brans further wrote that Mach’s Principle can be interpreted as requiring a theory in which the Newtonian inertial forces find their origin in the observed acceleration (rotation) of the distribution of the stars. Einstein then noted that according to the general theory of relativity, the mechanism providing such inertial effects has a gravitational nature. It is the inertial dragging associated with masses that are observed to rotate or have a translationally accelerated~motion.

In 1966, D. R. Brill and J. M. Cohen~\cite{B14} presented a calculation of the dragging angular velocity inside a massive shell with radius \emph{R}, Schwarzschild radius $r_{\rm s}$ and angular velocity $\omega$ beyond the weak field approximation. Their result is valid for arbitrarily strong gravitational fields although they restricted the calculation to slow rotation. This article represents a decisive step toward a resolution of the problem associated with the globally empty Minkowski spacetime. They showed that  to first order in \emph{$\omega$} spacetime inside the shell is~flat.

Let us define an inertial dragging coefficient as the ratio of the angular velocity, \emph{$\Omega$}, of an inertial frame inside the shell and the angular velocity of the shell, $K_{\rm IN}$ = $\Omega$/$\omega$. Brill and Cohen deduced that the inertial dragging coefficient, as expressed in isotropic coordinates,~is
      \begin{equation}
\label{eq:FD1}
K_{IN} = \frac{4r_{S}\left( {2R - r_{S}} \right)}{\left( {3R - r_{S}} \right)\left( {R + r_{S}} \right)}
\tag{1}
\end{equation}

We give a deduction of this formula in Appendix~\ref{app:app1}, providing more details than in the article~\cite{B14}.

If the shell has a radius equal to its Schwarzschild radius, $R = r_{\rm s}$ Equation (1) gives $K_{\rm IN} = 1$, i.e.,~$\Omega$ = $\omega$. This is called \emph{perfect inertial dragging}. Brill and Cohen note the Machian character of this result defining the contents of ``Machian" in the following way: ``For mass shells comprising more nearly \emph{all} the matter in the universe than those treated by Thirring, Mach’s principle suggests that the inertial properties of space inside the shell no longer depend on the inertial frames at infinity, but are completely determined by the shell itself". Having deduced the result above, they write that within the limit that the radius of the shell approaches its Schwarzschild radius, the rotation rate of the inertial frames inside the shell approaches the rotation rate of the shell. In other words, in this limit, \emph{the~inertial properties of space inside the shell no longer depend on the inertial frames at infinity, but are completely determined by the shell itself.}

They further write: ``A shell of matter of radius equal to its Schwarzschild radius has often been taken as an idealized cosmological model of our universe. Our result shows that in such a model there cannot be a rotation of the local inertial frame in the center relative to the large masses in the universe. In this sense, our result explains why the ``fixed stars" are indeed fixed in our inertial frame, and in this sense the result is consistent with Mach’s principle". The condition that the mass inside the cosmic horizon has a Schwarzschild radius equal to the horizon radius shall here be called \emph{the BC-condition for perfect inertial dragging}.

An exact solution of Einstein’s field equation, similar to that of Brill and Cohen where the restriction to slow rotation is not needed, will now be briefly discussed. In the Kerr spacetime, the~angular velocity of a particle with zero angular momentum (ZAM) is~\cite{B15}
      \begin{equation}
\label{eq:FD2}
\text{$\Omega$}_{Z} = \frac{2mac}{r^{3} + ra^{2} + 2ma^{2}}
\tag{2}
\end{equation}
      where $m = GM/c^2$ is the gravitational length of the central rotating body, and $a = J/Mc$ where $J$ is the angular momentum of the central mass (note that $a$ has dimension length).

In 1981, C. A. Lopez~\cite{B16} found a source of the Kerr spacetime. In 1985, \O . Gr\o n~\cite{B17} gave a much simpler deduction of this source and discussed some of its physical properties. The source is a shell with radius $r_{0} $ rotating with an angular~velocity
      \begin{equation}
\label{eq:FD3}
{\omega = \frac{ac}{a^{2} + r_{0}^{2}}}.
\tag{3}
\end{equation}

The radius of the exterior horizon in the Kerr metric~is
	  \begin{equation}
\label{eq:(4)}
r_{+} = m + \sqrt{m^{2} - a^{2}}
\tag{4}
\end{equation}

Hence, if the radius of the shell is equal to the horizon radius $r_{0} = r_{+} $, the ZAM angular velocity just outside the shell is equal to the angular velocity of the~shell,
	  \begin{equation}
\label{eq:(5)}
\text{$\Omega$}_{Z}(r_{+}) = \omega(r_{+}) = \frac{ac}{2mr_{+}}
\tag{5}
\end{equation}

Demanding continuity of the dragging angular velocity at the shell, it follows that the inertial frames in the Minkowski spacetime inside the shell are co-moving with the shell. There is perfect dragging of the inertial frames inside the shell. The properties of the shell, and of spacetime outside and inside the shell, solve Einstein’s field equations without needing the assumptions of weak fields and slow rotation. The inertial properties of space inside the shell, such as the Coriolis acceleration, do not depend on any property of an asymptotic far away region, only on the state of motion of the reference frame relative to the shell~\cite{B18}.

The phenomenon of perfect dragging has recently been demonstrated by C. Schmid~\cite{B19,B20,B21} in the context of the most general linear perturbations of the FLRW universe models, including perturbations that represent vorticity and rotational~motion.

\section{The Cosmic Causal Mass \label{sect:sec2}}

When we look outwards in space, we look backwards in time, because we see an object the way it was when it emitted the light that we receive. Remarkably, gravitational waves move at the velocity of light. Although it has a quantum mechanical explanation in the fact that both photons and gravitons are massless, it is a strange coincidence from a classical point of view, possibly indicating a deep connection between gravity and~electromagnetism.

It means that when we search for sources of gravitational effects that have propagated undisturbed from a changing source to an observer, neglecting tales of gravitational waves that can be contributions from the inside of the light cone, we must look at events along the past light cone. The gravitational effect that we study in this paper is inertial dragging, because it is relevant for the question of whether accelerated motion (translational and rotating) is relative according to the general theory of relativity. We search for cosmic sources of inertial dragging here and now. Hence, we introduce in this paper the concept \emph{causal mass}, i.e., the mass which produces gravitational effects here and now. When the causal mass at the point of time $t_{0} $ of an observer at \emph{r} = 0, is calculated by performing an integral with a mass element formed as a spherical shell about the observer with coordinate radius and thickness \emph{r} and \emph{dr}, respectively, the mass of the element is calculated by inserting the value of the density at the emission time of the considered mass element on the past light~cone.

Motivated by the papers mentioned above, we shall here calculate the mass along the past light cone within the cosmic particle horizon of an observer. This is the mass that acts causally upon the observer at the present time. The cited papers indicate that there will be at least close to perfect inertial dragging in our universe if the Schwarzschild radius of the causal mass inside the horizon is equal to or greater than the radius of the particle~horizon.

There are two main intentions of this paper. Firstly, we wanted to develop a better understanding of the concept ‘rotating universe model’ by studying the G\"{o}del universe. In particular, we use the concept of ‘perfect inertial dragging’ introduced by Brill and Cohen~\cite{B14} to obtain a better understanding of the rotation of the G\"{o}del universe. Secondly, we wanted to improve the study presented by Gr\o n and Jemterud~\cite{B22}, where they investigated the BC-condition for perfect inertial dragging by calculating the present amount of mass inside the cosmic horizon. Realizing that gravitational effects are not instantaneous, but move with the velocity of light according to Einstein’s theory, we wanted to conduct a more realistic investigation of whether the BC-condition for perfect inertial dragging is fulfilled in our universe, by calculating the mass along the past light cone inside the cosmic~horizon.

\subsection{The Causal Mass of the Einstein--de Sitter Universe \label{sect:sec2dot1}}

In order to illustrate our main result by a simple example, permitting analytical expressions in terms of elementary functions, we shall first consider the Einstein--de Sitter universe. This is a flat universe containing only dust. It has scale~factor
        \begin{equation}
\label{eq:FD4}
{a\left( t \right) = \left( \frac{t}{t_{0}} \right)^{2/3}},
\tag{6}
\end{equation}
        where $t_{\rm 0}$ is the present age of the universe, and the scale factor has been normalized to $\alpha(t_{\rm 0}) = 1$. The~Hubble age is $t_{\rm H} = 1/H_{\rm 0} $, where $H_{\rm 0} $ is the present value of the Hubble parameter. Inserting the most recent Planck value of the Hubble parameter gives $t_{\rm H} $ = 13.9 Gy~\cite{B23}. The age of this universe is
        \begin{equation}
\label{eq:FD5}
{t_{0ED} = \left( {2/3} \right)\, t_{H}}.
\tag{7}
\end{equation}

The physical radius of the particle horizon is (we are using units so that $c$ = 1),
        \vspace{12pt}
\begin{equation}
\label{eq:FD6}
{R_{PH}\left( t \right) = a\left( t \right){\int\limits_{0}^{t}{\frac{1}{a\left( t \right)}dt}}}.
\tag{8}
\end{equation}

Inserting the scale factor (6) gives the horizon radius of the Einstein--de Sitter~universe,
        \begin{equation}
\label{eq:FD7}
{R_{PHED}\left( t \right) = 3\, t_{0}^{2/3}t^{1/3}}.
\tag{9}
\end{equation}

The present radius~is
        \begin{equation}
\label{eq:FD8}
{R_{PHED}\left( t_{0} \right) = 3\, t_{0ED} = 2t_{H}}.
\tag{10}
\end{equation}

Hence, the present radius of the particle horizon is 27.8 Gly in an Einstein--de Sitter universe with the measured value of the Hubble~parameter.

We shall now calculate the Schwarzschild radius of the causal mass inside the particle horizon. It~is calculated by integrating along the past light cone, i.e., the density is evaluated at retarded points of~time,
        \begin{equation}
\label{eq:FD9}
r_{SED} = 2GM = 8\pi G{\int\limits_{0}^{r_{0PH}}{\left\lbrack {{\rho}\sqrt{g}} \right\rbrack_{t - r}dr =}}8\pi G{\int\limits_{0}^{r_{0PH}}{\left\lbrack {{\rho} a^{3}} \right\rbrack_{t - r}r^{2}dr}}
\tag{11}
\end{equation}

Here, \emph{g} is the determinant of the spatial part of the metric, and $r_{\rm 0PH}$ is the present radius of the particle horizon. Since the density of the dust is $\rho = \rho_0\alpha^{-3}$, we get
        \begin{equation}
\label{eq:FD10}
r_{SED} = \left( {8\pi G/3} \right){\rho}_{0}r_{0PH}^{3}
\tag{12}
\end{equation}

Using Equation (10) and given that the present density is equal to the present value of the critical~density,
        \begin{equation}
\label{eq:FD11}
{\rho}_{0} = \frac{3H_{0}^{2}}{8\pi G} = \frac{3}{8\pi Gt_{H}^{2}}
\tag{13}
\end{equation}

We~get
        \begin{equation}
\label{eq:FD12}
r_{SED} = 8t_{H}
\tag{14}
\end{equation}

Inserting Planck data gives $r_{SED} $ = 108.8 Gly. We see that for this universe model, $r_{SED} = 4R_{PHED} $. This indicates that there is perfect inertial dragging in this~universe.

The look-back distance is the radius of a surface S around an observer equal to the velocity of light times the age of the universe, $r_{LED} = t_{0} $. Inserting this as the upper limit in the integral (7), we find that the Schwarzschild radius of the mass inside S is equal to the look-back distance, $r_{SHED} = r_{LED} $. This corresponds to the condition for perfect inertial dragging used in~\cite{B22}. However, from a causal point of view, the relevant surface is the particle~horizon.

\section{Cosmic Inertial Dragging \label{sect:sec3}}

\subsection{Non-Rotating Universe \label{sect:sec3dot1}}

We shall here investigate the condition for perfect inertial dragging in the FLRW-universe models. In order to do so, we shall first describe the non-rotating universe models from a rotating reference frame, for example, the rest frame of the Earth. Hence, we introduce co-moving coordinates in a rotating frame by making the coordinate~transformation
        \begin{equation}
\label{eq:FD13}
{{\phi}\prime = {\phi} - \omega\, t},
\tag{15}
\end{equation}
        keeping the other coordinates unchanged. Inserting this into
        \begin{equation}
\label{eq:FD14}
{ds^{2} = - dt^{2} + a^{2}\left( t \right)\left( {dr^{2} + r^{2}d\theta^{2} + r^{2}\sin ^{2}\theta\, d{\phi}^{2}} \right)},
\tag{16}
\end{equation}
        and considering the equatorial plane with \emph{$\theta$} = \emph{$\pi$}/2 gives the line element
        \begin{equation}
\label{eq:FD15}
{ds^{2} = - \left( {1 - a^{2}r^{2}\omega^{2}} \right)dt^{2} + 2a^{2}r^{2}\omega\, d{\phi}\prime dt + a^{2}\left( {dr^{2} + r^{2}d{\phi}\prime^{2}} \right)}.
\tag{17}
\end{equation}

The Lagrange function per unit mass of a free particle moving along a circular path with constant radius~is
        \begin{equation}
\label{eq:FD16}
{L = - \frac{1}{2}\left( {1 - a^{2}r^{2}\omega^{2}} \right){\overset{.}{t}}^{2} + a^{2}r^{2}\omega\,\overset{.}{t}\,\overset{.}{{\phi}}\prime + \frac{1}{2}a^{2}r^{2}\overset{.}{{\phi}}\prime^{2}}.
\tag{18}
\end{equation}

The conserved angular momentum per unit mass~is
        \begin{equation}
\label{eq:FD17}
{p_{{\phi}} = \frac{\partial L}{\partial\overset{.}{{\phi}}\prime} = a^{2}r^{2}\left( {\omega\,\overset{.}{t} + \overset{.}{{\phi}}\prime} \right)}.
\tag{19}
\end{equation}

A particle with zero angular momentum (ZAM), with $p_{{\phi}} = 0 $, has a coordinate angular~velocity
        \begin{equation}
\label{eq:FD18}
{\omega_{ZAM} = \frac{d{\phi}\prime}{dt} = \frac{\overset{.}{{\phi}}\prime}{\overset{.}{t}} = - \omega}.
\tag{20}
\end{equation}

This means that the ZAM particle rotates together with the reference particles of the universe. Hence, there is perfect inertial dragging independent of the content of the universe, even for the empty Milne universe. The reason is that we have considered a non-rotating~universe.

The meaning of the term ‘rotating universe’ is that the reference particles of the universe, \mbox{i.e., the free} particles defining the 3-space, corresponding to clusters of galaxies in our universe, rotate relative to an inertial compass, for example, the swinging plane of a Foucault pendulum. In~a~non-rotating universe, the swinging plane of a Foucault pendulum is not rotating relative to the reference particles of the universe. This means that \emph{there is perfect inertial dragging in a non-rotating universe}. That is the explanation of the result shown in Equation (20).

However there remains one problem. In the general theory of relativity the significance of the Minkowski spacetime is that it is used as the asymptotic metric outside a localized mass distribution, for example in the Kerr spacetime. This means that absolute rotation is introduced into the general theory of relativity through the choice of boundary condition that the Minkowski spacetime is globally empty when solving Einstein’s field~equations.

In connection with Mach’s principle it is natural to impose the boundary condition that the asymptotically empty Minkowski spacetime is replaced by a Minkowski universe with a mass shell at infinity with radius equal to its own Schwarzschild radius. Since the Milne universe is the same as the Minkowski universe as described from an expanding reference frame, the same applies to the Milne~universe.

Then there will be a causal perfect inertial dragging in the Milne universe. Hence this new boundary condition is necessary to save the consistency of the relativity of rotational motion~\cite{B18}.

\subsection{The Rotating G\"{o}del Universe \label{sect:sec3dot2}}

Let us consider the G\"{o}del universe. The 3-space has a cylindrical symmetry. The kinematic properties of this universe have recently been vividly illustrated by Buser, Karjari and Schleich~\cite{B24}. They employed the technique of ray tracing and visualized various scenarios to bring out the optical effects experienced by an observer located in this universe. In cylindrical coordinates, the line-element takes the form~\cite{B25}
        \begin{equation}
\label{eq:FD19}
ds^{2} = 4a^{2}\left\lbrack {- dt^{2} + dr^{2} + dz^{2} + \left( {\text{sinh}^{2}r - \text{sinh}^{4}r} \right)d{\phi}^{2} - 2\sqrt{2}\text{sinh}^{2}r\, d{\phi}\, dt} \right\rbrack
\tag{21}
\end{equation}

The coordinates are co-moving with a set of free particles that act as reference particles for the 3-space. The angular momentum per unit mass of a free particle moving along a circular path with $dr = dz = 0 $~is
        \begin{equation}
\label{eq:FD20}
{p_{{\phi}} = 4a^{2}\left( {\text{sinh}^{2}r - \text{sinh}^{4}r} \right)\overset{.}{{\phi}} - 4a^{2}\sqrt{2}\text{sinh}^{2}r\text{  }\overset{.}{t}}.
\tag{22}
\end{equation}

Hence, a ZAM particle i.e., a particle with $p_\phi = 0$, has a coordinate angular~velocity
        \begin{equation}
\label{eq:FD21}
{\omega_{ZAM} = - \dfrac{\sqrt{2}}{1 - \text{sinh}^{2}r}}.
\tag{23}
\end{equation}

In the Kerr spacetime, the angular velocity calculated in this manner is said to represent the inertial dragging~effect.

In descriptions of rotating universe models, it has not been normal to write about inertial dragging. One has instead considered the vorticity tensor of the 4-velocity field of the reference particles. In~general, the covariant components of the vorticity tensor are defined~by
        \begin{equation}
\label{eq:FD22}
{\omega_{\mu\nu} = u_{\lbrack{\mu;\nu}\rbrack} + {\overset{.}{u}}_{\lbrack\mu}u_{\nu\rbrack}},
\tag{24}
\end{equation}
        where the bracket means antisymmetrization. The first term is the curl of the 4-velocity field, and the second term the antisymmetrized 4-acceleration. Thus, the last term vanishes for free particles.

The coordinates are co-moving with the reference particles. It follows from the line element (21) that $\overset{.}{t} = 1/2a $ for a particle permanently at rest in the coordinate system. Hence, their 4-velocity~is
        \begin{equation}
\label{eq:FD23}
{\mathbf{u} = u^{t}\mathbf{e}_{t} = \overset{.}{t}\mathbf{e}_{t} = \left( {1/2a} \right)\mathbf{e}_{t}}.
\tag{25}
\end{equation}

The covariant components of this 4-velocity field are given by $u_{\mu} = g_{\mu\nu}u^{\nu} $ which gives the non-vanishing~components
        \begin{equation}
\label{eq:FD24}
{u_{t} = - \, 2a\text{      },\text{      }u_{{\phi}} = \sqrt{2}\, a\text{  }\text{sinh}^{2}r}.
\tag{26}
\end{equation}

Due to the symmetries of the G\"{o}del universe, it is clear that the 4-acceleration of the reference particles has vanishing $t - , $ ${\phi} - , $ and $z - $components. The \emph{r}-component is $a^{r} = u^{r}{}_{;\nu}u^{\nu} = u^{r}{}_{;t}u^{t} = \text{$\Gamma$}^{r}{}_{tt}{\overset{.}{t}}^{2} $. Calculating $\text{$\Gamma$}^{r}{}_{tt} $ from the line element (21), one finds $\text{$\Gamma$}^{r}{}_{tt} = 0 $ Hence, the 4-acceleration of these particles vanish, meaning that the reference particles of the G\"{o}del universe are freely~falling.

This means that the expression of the covariant components of the vorticity tensor of the 4-velocity field of the reference particles reduces~to
        \begin{equation}
\label{eq:FD25}
\omega_{\mu\nu} = \frac{1}{2}\left( {u_{\mu,\nu} - u_{\nu,\mu}} \right)
\tag{27}
\end{equation}
        where the covariant derivatives are replaced by ordinary derivatives due to the symmetry of the Christoffel symbols. The only non-vanishing components are
        \begin{equation}
\label{eq:FD26}
{\omega_{{\phi} r} = - \omega_{r{\phi}} = \sqrt{2}\, a\text{sinh}r\,\cosh r}.
\tag{28}
\end{equation}

The contravariant components of the vorticity tensor are given~by
        \begin{equation}
\label{eq:FD27}
{\omega^{\mu\nu} = g^{\mu\alpha}g^{\mu\beta}\omega_{\alpha\beta} = 2g^{\mu r}g^{\nu{\phi}}\omega_{r{\phi}}}.
\tag{29}
\end{equation}

The rotation scalar is defined~by
        \begin{equation}
\label{eq:FD28}
{\text{$\Omega$} = \sqrt{\left( {1/2} \right)\omega^{\mu\nu}\omega_{\mu\nu}}}.
\tag{30}
\end{equation}

We then~have
        \begin{equation}
\label{eq:FD29}
\omega^{\mu\nu}\omega_{\mu\nu} = 2g^{rr}g^{{\phi}{\phi}}\omega_{r{\phi}}^{2}
\tag{31}
\end{equation}

The contravariant components of the metric tensor are defined by $g^{\mu\alpha}g_{\alpha\nu} = \delta^{\mu}{}_{\nu} $ which here leads~to
        \begin{equation}
\label{eq:FD30}
{g^{rr} = \frac{1}{g_{rr}} = \frac{1}{4a^{2}}\text{      },\text{      }g^{{\phi}{\phi}} = \frac{- g_{tt}}{g_{t{\phi}}^{2} - g_{tt}g_{{\phi}{\phi}}} = \frac{1}{4a^{2}\text{sinh}^{2}r\cosh ^{2}r}},
\tag{32}
\end{equation}
        giving
        \begin{equation}
\label{eq:FD31}
{\text{$\Omega$} = \frac{1}{\sqrt{2}a}}.
\tag{33}
\end{equation}

Hence, the local angular velocity in a plane orthogonal to the symmetry axis of the ‘river of space’~\cite{B26} in the G\"{o}del spacetime has a constant value. This means that the swinging plane of a Foucault pendulum rotates with this angular velocity with respect to the reference particles defining the space of the G\"{o}del~universe.

It should be noted that, \emph{in this universe, the cosmic mass does not cause perfect inertial dragging}. The~swinging plane of a Foucault pendulum is not rotating together with the cosmic masses in this~universe.

Ozsv\'ath and Sch\"ucking~\cite{B27} have discussed Mach’s principle in connection with a finite rotating universe model. This solution was said to have a non-Machian character. However, the phenomenon of inertial dragging was not mentioned. Thus, it was not noted that the rotation of the universe means that there is not perfect inertial dragging in such a~universe.

The connection between the perfect inertial dragging and vanishing rotation of the universe means that the existence of rotating universe models need not be anti-Machan after all. It only means that there is not a sufficiently large causal mass in the universe to drag the swinging plane of a pendulum around together with the cosmic~mass.

Saadeh and coworkers~\cite{B28} have recently applied a new upper limit to how fast our universe can rotate which is about ten times tighter than previous limits. Using background temperature and polarization data from the Planck satellite, they found that the angle at which the universe could have rotated during its Hubble age fulfils $\omega\, t_{H} < 5.2 \cdot 10^{- 11} $ radians. Hence, for all practical purposes, one may say that our universe is non-rotating. We therefore expect that there is sufficient causal mass in the universe to cause perfect inertial dragging in our~universe.

\subsection{Inertial Dragging in a Cosmological Context \label{sect:sec3dot3}}

The work of Brill and Cohen~\cite{B14} has been followed up. In a paper with the title Coriolis Effects in the Einstein Universe, A. Lausberg~\cite{B29} has studied inertial effects induced by the slow rotation of a shell with finite thickness in a closed universe filled with matter, i.e., he has perturbed Einstein’s static universe. He has allowed part of the Einstein universe between two radii $r_{0} $ and $r_{1} $ to rotate rigidly with an angular velocity $\omega_{S} $ so that $r_{1}\omega_{S} < < 1 $. We will here assume that the rotating part of the universe covers the space from the observer at $r_{0} $ = 0 to a radius \emph{r}. Lausberg found that the ratio of the mass in the rotating part of the universe and the total mass~is
        \begin{equation}
\label{eq:FD32}
M_{S}/M = \left( {1/\pi} \right)\left( {r - \sin r\cos r} \right)
\tag{34}
\end{equation}
        where $r \in \left\lbrack {0,\,\pi} \right\rbrack $. Let $\omega_{0} $ be the angular velocity of an inertial compass at $r_{0} $ = 0. Lausberg called the ratio $\omega_{0}/\omega_{S} $ the dragging coefficient, and found that in the present universe model it has a value (correcting a sign error)
        \begin{equation}
\label{eq:FD33}
{\omega_{S}/\omega_{0} = \left( {1/3\pi} \right)\left( {r - 3\sin r\cos r + 2\pi\sin ^{2}r + 2r\cos ^{2}r} \right)}.
\tag{35}
\end{equation}

\begin{figure}[tbp]
\begin{center}
\includegraphics[width=17cm]{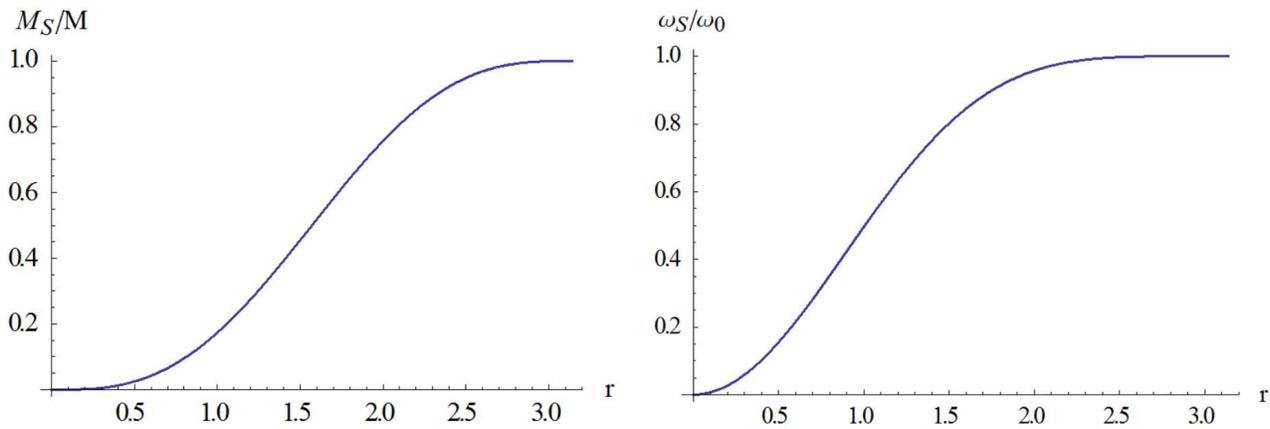}
\caption{The mass ratio and the dragging coefficient in a perturbed Einstein universe, where the part of the universe out to a radius \emph{r} is rotating~slowly.}
\label{fig:fig1}
\end{center}
\end{figure}

\begin{figure}[ptb]
\begin{center}
\includegraphics[width=7cm]{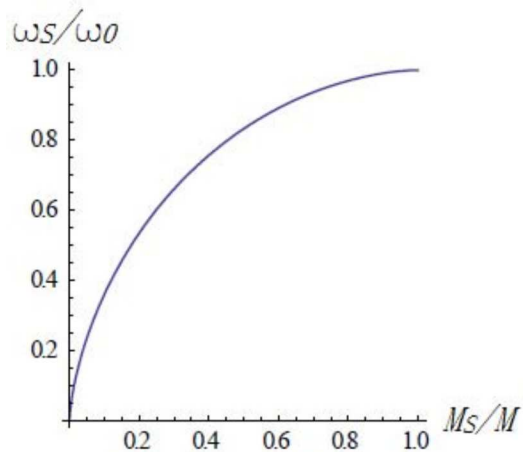}
\caption{The inertial dragging coefficient as a function of the mass ratio in a rotationally perturbed Einstein universe. There is perfect inertial dragging, meaning that the inertial compass at \emph{r} = 0 follows the cosmic mass, if all of the matter in the universe~rotates.}
\label{fig:fig2}
\end{center}
\end{figure}

It represents the dragging effect of the rotating cosmic mass upon, for example, the swinging plane of a Foucault pendulum at the origin. The mass ratio and dragging coefficient are shown graphically as functions of the radius in Figure~\ref{fig:fig1}, and the dragging coefficient is shown as a function of the mass ratio in Figure~\ref{fig:fig2} (using ParametricPlot in Mathematica).

Lindblom and Brill~\cite{B30} examined inertial dragging due to a freely falling, slowly rotating, shell of dust in a dust-dominated FLRW universe. They found that for a shell falling from an initial very large radius, the inertial dragging~coefficient
        \begin{equation}
\label{eq:FD34}
K_{IN} = \left\lbrack {1 + \frac{6R^{2}\left( {4R - r_{S}} \right)^{2}}{r_{S}\left\lbrack {64R^{3} - r_{S}\left( {4R + r_{S}} \right)^{2}} \right\rbrack}} \right\rbrack^{- 1}
\tag{36}
\end{equation}

Here, the initial value of the dragging coefficient is $\underset{R\rightarrow\infty}{\text{lim}}K_{IN} = 0 $, and the final value when the radius of the shell approaches the horizon radius is $K_{IN}\left( {R = r_{S}/4} \right) = 1 $. Generally, the dragging angular velocity is determined both by the rotating mass of the shell and the non-rotating mass outside it. The authors note that this result has a Machian~character.

In 1980, Lewis published a paper~\cite{B31} where he, too, discussed inertial dragging in the framework of an expanding, spatially closed FLRW universe model. He discussed a model with a rotating shell having Minkowski spacetime inside the shell, an expanding space outside it, but the shell was not a freely falling dust shell. He argued that there were inertial dragging effects in accordance with the spirit of Mach’s~principle.

Klein~\cite{B32} published an interesting paper in 1993, in which he discussed rotational perturbations and inertial dragging in an expanding dust-dominated FLRW universe. He also perturbed the FLRW model by introducing a slowly rotating dust shell, this time expanding together with the dust of the RLRW-model. It follows from the Friedmann equation $\kappa\rho = 3H^2$, where $\rho$ is the density of the dust and $H$ the Hubble parameter, that the dust and the shell vanish in the case that there is no expansion. Hence, Klein’s model has no stationary~limit.

Klein found that the inertial dragging coefficient~is
        \begin{equation}
\label{eq:FD35}
K_{IN} = \left\lbrack {1 + \frac{3}{2R^{2}}\left( {\frac{{\overset{.}{a}}^{2}}{1 + \alpha} - \frac{a\overset{..}{a}}{\alpha}} \right)^{- 1}} \right\rbrack^{- 1}\text{   },\text{      }\alpha = \sqrt{1 + R^{2}{\overset{.}{a}}^{2}}
\tag{37}
\end{equation}

For a dust-dominated universe the Friedmann equations imply that $2a\overset{..}{a} + {\overset{.}{a}}^{2} = 0 $ and ${\rho} a^{3} = {\rho}_{0} $. Using these equations in Equation (37) leads~to
        \begin{equation}
\label{eq:FD36}
K_{IN} = \left\lbrack {1 + \frac{9\alpha\left( {1 + \alpha} \right)a\left( t \right)}{\kappa{\rho}_{0}R^{2}\left( {1 + 3\alpha} \right)}} \right\rbrack^{- 1}
\tag{38}
\end{equation}

Considering, for example, a flat dust-dominated universe with scale factor (6) so that \emph{a}(0) = 0, the~initial value of the dragging coefficient is $\left. K_{IN}\left( 0 \right)\rightarrow 1 \right. $. In the far future, \emph{a}$\rightarrow$$\infty$ and $\left. K_{IN}\rightarrow 0 \right. $. Klein gives the following comment. The fact that $\left. K_{IN}\rightarrow 0 \right. $ in the far future is expected from a Machian point of view: In an expanding universe, the matter density decreases as a function of time together with the mass of the shell, which has been assumed to be equal to the mass of the dust inside the shell in the unperturbed FLRW model. Thus, the influence of the matter on the inertial frames described by the dragging coefficient will~vanish.

He comments on the dependence of $K_{IN} $ upon the radius \emph{R} of the shell as follows. If \emph{R} is getting large (but the rotation is still so slow that \emph{R$\omega$} \textless{}\textless{} 1.), the dragging coefficient $\left. K_{IN}\rightarrow 1 \right. $, i.e., perfect dragging will be approached. In this case, nearly the whole matter of the universe is distributed on the shell, and therefore the inertial dragging becomes~perfect.

Further progress came in 1995 with a work by Lynden-Bell, Katz and Bi\v c\'ak~\cite{B33}. They considered a system of freely falling spheres rotating about a common axis as an expanding universe model, and~found in the case of a closed universe that the dragging of inertial frames is determined uniquely by the matter~distribution.

Then, in the year 2000, Dole\v zel, Bi\v c\'ak and Deruelle~\cite{B34} followed up with an interesting article in which they investigated a spacetime consisting of an empty void separated from a dust-dominated FLRW universe by a spherically symmetric, slowly rotating shell, co-moving radially with the cosmic dust. This work was reviewed and extended by Doležel in 2002~\cite{B35}.

They start by making the following formulation of Mach’s principle, which is also the version favored by the present authors: ``Mach’s principle is the proposition that the motion of local inertial frames is determined by an ‘average’ motion of the matter in the universe". The mechanism making the realization of this principle possible, according to the general theory of relativity, is the phenomenon of inertial dragging, in particular \emph{perfect inertial dragging,} which means that the dragging due to the cosmic masses is so strong that a local inertial frame has no acceleration or rotation relative to the large-scale distribution of the mass in the~universe.

Their main result is the calculation of the inertial dragging coefficient inside an empty void in an expanding universe. It follows from Equations (19), (30) and (39) of Dole\v zel, Bi\v c\'ak and Deruelle together with the Friedmann equation $2a\overset{..}{a} + {\overset{.}{a}}^{2} = 0 $ for a flat, dust-dominated universe~that
        \begin{equation}
\label{eq:FD37}
K_{IN} = \frac{2 + 3R^{2}{\overset{.}{a}}^{2} - 2\alpha}{3\alpha}
\tag{39}
\end{equation}

Here, $R $ is the constant co-moving radial coordinate of the shell at the boundary between the expanding void and the dust-dominated universe outside, and $\alpha $ is given in Equation (37). This~equation has been deduced under the assumption of slow rotation. Perfect inertial dragging, $K_{IN} = 1 $, requires $R\overset{.}{a} = \sqrt{\left( {13 + 5\sqrt{37}} \right)/18} \approx 1.5 $. In order to have perfect inertial dragging at the present time with $\overset{.}{a} = 2/3t_{0} $, the coordinate radius of the void must be $R \approx 2.3\, t_{0} $. For $R\overset{.}{a} > > 1 $, we see that $\left. \alpha\rightarrow r\overset{.}{a} \right. $ and $\left. K_{IN}\rightarrow\left( {1/3} \right)R\overset{.}{a} > 1 \right. $ which means that the dragging angular velocity is larger than that of the rotating boundary of the void. This is obviously unphysical. However, in the model of Dole\v zel, Bi\v c\'ak and Deruelle, the proper angular velocity of the shell is approximately $\omega_{S} \propto \left\lbrack {R^{5}a^{5}\left( {\overset{.}{a}/a} \right)^{\text{$\cdotp $}}} \right\rbrack^{- 1} $ when $R\overset{.}{a} > > 1 $. Thus, the assumption of slow rotation, $Ra\omega_{S} < < 1 $  requires that the quantity $\left| {a^{4}\left( {\overset{.}{a}/a} \right)^{\text{$\cdotp $}}} \right| $ is sufficiently large. For a flat, dust-dominated universe with scale factor $a\left( t \right) = \left( {t/t_{0}} \right)^{2/3} $, we have $\left| {a^{4}\left( {\overset{.}{a}/a} \right)^{\text{$\cdotp $}}} \right| = \left( {2/3} \right)t_{0}^{- \, 8/3}t^{2/3} $. Hence, the expression (35) is not valid for very small values of $t $. It is only valid for $R\overset{.}{a} < 1 $ $R\overset{.}{a} < 1 $, and then the dragging coefficient is smaller than one. For large values of the cosmic time, $\left. \overset{.}{a}\rightarrow 0 \right. $ and $\left. \alpha\rightarrow 1 \right. $. Then, $\left. K_{IN}\rightarrow 0 \right. $ and the inertial dragging~vanishes.

\section{The Causal Mass inside the Particle Horizon in the Flat $\Lambda$CDM Universe \label{sect:sec4}}

We now consider the flat $\Lambda$CDM universe which has scale factor~\cite{B36}
      \begin{equation}
\label{eq:FD38}
{a\left( t \right) = A^{1/3}\text{sinh}^{2/3}\left( {t/t_{\text{$\Lambda$}}} \right),\text{      }A = \frac{1 - \text{$\Omega$}_{\text{$\Lambda$}}}{\text{$\Omega$}_{\text{$\Lambda$}}},\text{      }t_{\text{$\Lambda$}} = \frac{2}{\sqrt{3\text{$\Lambda$}}} = \frac{2\, t_{H}}{3\sqrt{\text{$\Omega$}_{\text{$\Lambda$}0}}}}.
\tag{40}
\end{equation}

In this case, the integrals (8) and (11) must be performed numerically. The present radius of the particle horizon in this universe~is
      \begin{equation}
\label{eq:FD39}
{r_{0PH} = r_{PH}\left( t_{0} \right) = \frac{2}{3}\frac{t_{H}}{\text{$\Omega$}_{\text{$\Lambda$}0}^{1/6}\text{$\Omega$}_{M0}^{1/3}}{\int\limits_{0}^{\text{arsinh}\sqrt{\text{$\Omega$}_{\text{$\Lambda$}0}/\text{$\Omega$}_{M0}}}\frac{dx}{\text{sinh}^{2/3}\left( x \right)}}}.
\tag{41}
\end{equation}

Inserting the Planck values $\text{$\Omega$}_{M0} = 0.32\,,\text{   }\text{$\Omega$}_{\text{$\Lambda$}0} = 0.68 $ and $t_{H} = 13.9\text{  }\text{Gy} $~\cite{B23}, and calculating the integral numerically gives $r_{0PH} = 45\text{  }\text{Gly} $.

The mass that acts causally upon the observer located at $r = 0 $ at the present time $t_{0} $ in the $\Lambda$CDM universe can be written as the~integral
      \begin{equation}
\label{eq:FD40}
M = 4\pi\int\limits_{0}^{r_{0ph}}{\lbrack{{\rho}_{M0}\  + \ {\rho}_{\Lambda}{({a\left( {t_{e}\left( r \right)} \right)^{3}})}}\rbrack}\ r^{2}dr.
\tag{42}
\end{equation}

Here, ${\rho}_{M0} $ and ${\rho}_{\Lambda} $ denote the present density of the dust and the vacuum, respectively, and $t_{e} $ denotes the emission time of a signal emitted at the coordinate distance \emph{r} and received at $r = 0 $ at time $t_{0} $. Accordingly, $t_{e} $ is a function of the coordinate distance \emph{r} the signal travels and is given implicitly by the~relation
      \begin{equation}
\label{eq:FD41}
r\left( t_{e} \right) = \int\limits_{t_{e}}^{t_{0}}\frac{c}{a\left( t \right)}dt,
\tag{43}
\end{equation}
      which may readily be rewritten as the first-order ordinary differential equation
      \begin{equation}
\label{eq:FD42}
\frac{dr}{dt} = \frac{c}{a\left( \left| t \right| \right)}
\tag{44}
\end{equation}
      with the initial condition r($- t_{0} $) = 0. The present Schwarzschild radius of this mass is $R_{S} = 2GM $.

Using the MATLAB ode45-solver with the built-in ‘event location’ capability, the coupled set of Equations (42) and (44) are solved numerically by the following procedure. For each discrete value $r_{n} $ of the independent variable \emph{r} used in evaluating the mass integral in Equation (42), the corresponding value of $t_{e} $ is obtained by solving Equation (44) combined with the event location feature to stop the solution at the specified value $r = r_{n} $. The value of the scale factor $a(t_{e}\left( r_{n} \right)) $ appearing in the integrand in Equation (42) is then evaluated at the corresponding emission time $t_{e}\left( r_{n} \right) $. Thus, we obtain the mass along the past light cone within the particle~horizon.

\begin{figure}[ptb]
\begin{center}
\includegraphics[width=15cm]{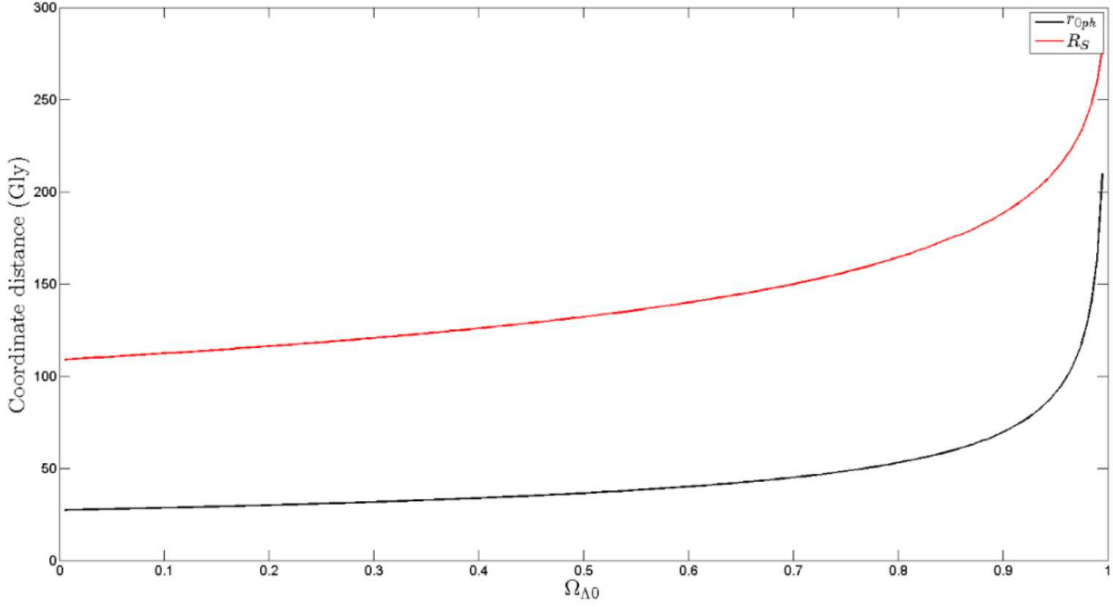}
\caption{The upper curve represents the Schwarzschild radius of the causal mass inside the present particle horizon in a flat $\Lambda$CDM universe, and the lower curve represents the present radius of the particle~horizon.}
\label{fig:fig3}
\end{center}
\end{figure}

As was noted after Equation (14), the Schwarzschild radius of the causal mass inside the present particle horizon is four times the radius of the horizon for the case that $\text{$\Omega$}_{\text{$\Lambda$}0} = 0 $, as is also seen from Figure~\ref{fig:fig3}. The Schwarzschild radius of the causal mass inside the horizon is larger than the radius of the horizon for all values of $\text{$\Omega$}_{\text{$\Lambda$}0} $. This indicates that there is perfect inertial dragging in our universe, and hence explains that the compass of inertia has a fixed orientation relative to the starry~sky.

\subsection{Contribution of Dust to the Causal Mass inside the Particle Horizon in the Flat $\Lambda$CDM Universe \label{sect:sec4dot1}}

In Equation (42), the mass inside the retarded particle horizon is given in terms of the present radius of the particle horizon. We will here consider the contribution of the dust and calculate the ratio of the Schwarzschild radius of the dust inside the retarded particle horizon and the present radius of the particle horizon for the $\Lambda$CDM-model. The first term gives the mass of the dust inside the retarded particle~horizon
        \begin{equation}
\label{eq:FD43}
M_{D} = \frac{4\pi}{3}{\rho}_{D0}r_{ph0}^{3}
\tag{45}
\end{equation}

The ratio of the Schwarzschild radius of the dust and the present radius of the particle horizon~is
        \begin{equation}
\label{eq:FD44}
{\frac{r_{SD0}}{r_{ph0}} = \frac{8\pi}{3}G{\rho}_{D0}r_{ph0}^{2}}.
\tag{46}
\end{equation}

The density parameter of the dust~is
        \begin{equation}
\label{eq:FD45}
{\text{$\Omega$}_{D0} = \frac{{\rho}_{D0}}{{\rho}_{cr}} = \frac{8\pi G{\rho}_{D0}}{3H^{2}}}.
\tag{47}
\end{equation}

Hence
        \begin{equation}
\label{eq:FD46}
{\frac{r_{SD0}}{r_{ph0}} = H^{2}\text{$\Omega$}_{D0}r_{ph0}^{2}}.
\tag{48}
\end{equation}

For this universe model, the Hubble age $t_{H} = 1/H $ is very close to the age, $t_{H} \approx 13,8Gy $. The~present radius of the particle horizon is $r_{ph0} = 45Gly $. Hence, we~get
        \begin{equation}
\label{eq:FD47}
{\frac{r_{SD0}}{r_{ph0}} = 3,3\text{$\Omega$}_{M0}}.
\tag{49}
\end{equation}

With $\text{$\Omega$}_{D0} $= 0.32 we~get
        \begin{equation}
\label{eq:FD48}
{\frac{r_{SD0}}{r_{ph0}} = 1.05}.
\tag{50}
\end{equation}

This shows that the dust alone gives perfect inertial~dragging.

As seen from Equation (42), the vacuum energy also contributes to the inertial dragging. This has been demonstrated previously in a different way by Farup and Gr\o n~\cite{B37}.

\section{On the Relativity of Rotation in Einstein’s Theory \label{sect:sec5}}

A hundred years ago, de Sitter published~\cite{B6} an article with this title. There he wrote: ``Rotation is thus relative in Einstein’s theory" He concludes his article with: ``For Einstein, who makes no difference between inertia and gravitation, and knows no absolute space, the accelerations which the classical mechanics ascribed to centrifugal forces are of exactly the same nature and require no more and no less explanation, than those which in classical mechanics are due to gravitational attraction".

De Sitter did not discuss the physical mechanism producing the centrifugal acceleration except mentioning that is has a gravitational origin. We now know that it is inertial dragging and that perfect inertial dragging is necessary in order for rotational motion to be relative. This means that the relativity of rotation according to Einstein’s theory is not a principle that is generally a consequence of Einstein’s theory, but it has a circumstantial nature. There exist universes where rotation is relative, and there also exist universes where this in not the~case.

We have shown that, in our universe, the causal mass inside the event horizon has a Schwarzschild radius which is larger than the horizon radions. Hence, the condition for perfect inertial dragging is fulfilled in our universe. This means that there are no centrifugal effects in a reference frame that is not rotating relative to an inertial compass in our universe. On the other hand, it also means that as observed on the Earth, the inertial compass rotates together with the average distribution of the cosmic mass. All of the centrifugal and Coriolis effects observed in this reference frame can then be explained as a gravitational effect of the rotating cosmic mass due to perfect inertial~dragging.

It is otherwise in the G\"{o}del universe. In this universe, it is not perfect inertial dragging. Hence, the cosmic mass rotates relative to an inertial compass in this universe. In the reference frame of the inertial compass, there are no centrifugal effects. Thus, in the rest frame of the cosmic mass, there are inertial effects. These cannot be fully explained as gravitational effects since there is not perfect inertial dragging. Hence, rotation is absolute in this universe since the centrifugal effects in the rest frame of the cosmic mass must then be explained as being due to a rotation of the cosmic mass, defining the reference frame with vanishing inertial effects as~non-rotating.

\section{Conclusions \label{sect:sec6}}

We have discussed the meaning of the concept ‘rotating universe’. The kinematic definition is that a universe is rotating if the 4-velocity field of the reference particles defining the 3-space of the universe has a non-vanishing vorticity. However, it is also normal to say that a universe is rotating if the reference particles, or the cosmic matter, rotate relatively to the fixed orientation of an inertial compass. In a non-rotating universe, an observer at rest on the Earth would see that, for example, the~swinging plane of a Foucault pendulum rotates together with the remote galaxies in the sky. According to the general theory of relativity, the physical mechanism connecting the motion of the swinging plane of a pendulum and the motion of the remote galaxies, is inertial dragging. There is perfect inertial dragging in a non-rotating universe, and in a rotating universe the inertial dragging is not perfect. Hence, in the G\"{o}del universe, for example, the cosmic inertial dragging is not~perfect.

In several works that we have reviewed above, it has been shown that inside a slowly rotating shell with mass so large that its Schwarzschild radius is equal to its radius, there is perfect inertial dragging inside the shell. In some of the models where this has been demonstrated, the mass of the shell is equal to the mass of the matter content of the universe model inside the shell. The principle of causality says that an event which is the cause of another event called the effect, cannot happen outside the past light cone of the effect. In this connection, we have introduced a new concept which we have called the causal mass of an observer. The relevance of the causality principle for the possible validity of the principle of relativity for rotational motion will now be~discussed.

Imagine an observer at an arbitrary point of a universe. She stands on a carousel which suddenly starts. At the moment that she observes that the distribution of stars on the heaven starts to rotate, she observes that centrifugal acceleration appears at her position. Assume that she considers herself at rest and that the stars suddenly start rotating. The rotation causes inertial dragging which seems to act instantaneously. At the same moment that she observes that the heaven starts rotating, she experiences centrifugal~acceleration.

This seems to require that gravity acts instantaneously, and due to the relativity of simultaneity another observer moving relative to the first may then experience the gravitational effect before the cause, in conflict with the principle of~causality.

This has been used as an argument that the observer is not permitted to regard herself as at rest and the universe to rotate around her, and thus that rotation is absolute in Einstein’s theory as well as in Newton’s~theory.

The situation does not, however, involve instantaneous action at a distance. The reason is that the far away masses did not start to rotate at the same moment as the observer saw it. The rotation started at an earlier time, equal to the time that light took to move from the mass point to the observer. Since gravitational waves are moving at the velocity of light, it follows that the gravitational effect of the sudden motion arrives at the same moment as she sees that motion. Hence, the principle of causality is obeyed, and the observer may indeed consider herself as at~rest.

Brill and Cohen argued, from an example with an observer inside a rotating shell, that a shell with mass so large that its Schwarzschild radius is equal to its radius, causes perfect inertial dragging inside the shell, which means that the swinging plane of a pendulum is dragged around together with the shell. This is necessary in order for an observer at the North Pole of the earth to be permitted to consider herself as at rest and the universe as rotating. Hence, the perfect inertial dragging is a requirement for the principle of relativity to be valid for rotating motion. We have called this the~BC-requirement.

Hence, it is reasonable to conjecture that there will be perfect inertial dragging in a universe if the Schwarzschild radius of the causal mass inside the particle horizon is equal to or greater than the radius of the particle horizon. This has motivated us to calculate the ratio of the Schwarzschild radius of the causal mass inside the particle horizon and the radius of this horizon in our universe, or at least in the favored model of our universe---the flat $\Lambda$CDM universe model. The result is shown in Figure~\ref{fig:fig3}. The Schwarzschild radius of the causal mass is more than three times the radius of the particle horizon. This indicates that there is perfect cosmic inertial dragging in our universe, in agreement with the results of Schmid~\cite{B19,B20,B21} based on different~methods.

Einstein’s general theory of relativity opens the possibility that accelerated and rotational motion is relative. It has not been proven that this is indeed the case. However, the mechanisms to make all motion relative are there. What is needed is the principle of equivalence and perfect inertial dragging. Our calculations show that the BC-condition for perfect inertial dragging due to the cosmic masses is satisfied in our universe. This means that Einstein’s extension of the principle of relativity to encompass, for example, rotating motion, is a consequence of the general theory of relativity as applied to our~universe.

If you stand on the North Pole of the Earth with a pendulum and a telescope, you will find what seems like a strange coincidence: The swinging plane of the pendulum follows the stars in the sky. According to Newton’s theory of gravitation, this only means that neither the swinging plane of the pendulum nor the distribution of stars rotate. It is the Earth that rotates. Einstein’s theory is more interesting. According to the general theory of relativity, this ‘coincidence’ indicates a deep connection between the swinging plane of the pendulum and the cosmic mass. They are connected gravitationally, and inertial dragging is the mechanism. Perfect inertial dragging from the cosmic masses makes the swinging plane of the pendulum follow the stars in the sky. In this way, the observer on the North Pole may indeed consider the Earth as non-rotating, and the swinging plane of the pendulum and the distribution of stars as rotating. This indicates that the principle of relativity for rotating motion may be valid in our universe according to Einstein’s~theory.

\vspace{6pt}
\small{\textbf{Acknowledgments:} We would like to thank all three referees for valuable suggestions that contributed significantly to improvements of the~article.}

\small{\textbf{Author Contributions:} Simen Br\ae ck performed the numerical calculation of the causal mass in the $\Lambda$CDM universe model and wrote the
Appendix~\ref{app:app1}. \O yvind Gr\o n wrote most of the main text and made significant conceptual contributions. Ivar Farup and Simen Br\ae ck carried out the calculations presented in the Appendix~\ref{app:app1}. All authors discussed the results and commented on the content of the~manuscript.}

\small{\textbf{Conflicts of Interest:} The authors declare no conflicts of~interest.}

\appendix

\makeatletter
\@addtoreset{table}{part}
\renewcommand{\thetable}{A\@arabic\c@table}
\setcounter{table}{0}
\@addtoreset{figure}{part}
\renewcommand{\thefigure}{A\@arabic\c@figure}
\setcounter{figure}{0}
\makeatother
\section{\noindent Detailed Calculation of the Brill--Cohen Formula for the Inertial Dragging inside a Slowly Rotating Shell \label{app:app1}}

We consider first an infinitely thin, nonrotating spherical shell of mass $m $ positioned at a coordinate radius $r = r_{0} $. In the regions inside the shell, there may be a spherical mass, $M_{-} $ at the center. Outside the shell, space is assumed to be empty. Jebsen-Birkhoff’s theorem~\cite{B38,B39} applied to the empty space inside and outside the spherically symmetric shell then implies that the spacetime geometry in both these regions is given by the Schwarzschild solution. Thus, in standard Schwarzschild coordinates $({\widetilde{t},\widetilde{r},\widetilde{\vartheta},\widetilde{{\varphi}}}) $, the spacetime geometry is described by the line~element
        \begin{equation}
\label{eq:FD49}
{ds^{2} = \left( {1 - \frac{2M_{\pm}}{\widetilde{r}}} \right)d{\widetilde{t}}^{2} - \frac{1}{1 - \frac{2M_{\pm}}{\widetilde{r}}}d{\widetilde{r}}^{2} + \ {\widetilde{r}}^{2}{({d{\widetilde{\vartheta}}^{2} + \sin ^{2}\widetilde{\vartheta}d{\widetilde{{\phi}}}^{2}})}},
\tag{A1}
\end{equation}
        where $M_{-} $ and $M_{+} $ denote constants of integration in the interior and exterior of the shell, respectively.

In this Appendix~\ref{app:app1}, we shall consider a spacetime where space is empty at the point $r = 0 $. Since the solution must be regular at this point, the constant $M_{-} $ must vanish. As a result, the geometry inside the shell is that of the flat Minkowski~spacetime.

Following Brill and Cohen~\cite{B14}, we shall hereafter adopt isotropic coordinates for the Schwarzschild geometry. In these coordinates, the shell has a radius R and a Schwarzschild radius $r_{S} $. Then, in isotropic coordinates, the line element for the spacetime outside and inside the shell can be written~as
        \begin{equation}
\label{eq:FD50}
ds^{2} = V\left( r \right)^{2}dt^{2} - \psi\left( r \right)^{4}\left( {dr^{2} + \ r^{2}\left( {d\vartheta^{2} + \sin ^{2}\vartheta d{\varphi}^{2}} \right)} \right)\ ,
\tag{A2}
\end{equation}
        where
        \begin{equation}
\label{eq:FD51}
V\left( r \right) = \left\{ {\begin{array}{c}
{\left( {r - r_{S} = \pi r^{2}} \right)/\left( {r + r_{S}} \right)\ \text{for}\ r > R} \\
{\left( {R - r_{S}} \right)/\left( {R + r_{S}} \right) \equiv V_{0}\ \text{for}\ r < R} \\
\end{array}\ ,} \right.
\tag{A3}
\end{equation}
        and
        \begin{equation}
\label{eq:FD52}
\psi\left( r \right) = \left\{ \begin{array}{c}
{1 + r_{S}/r\ \text{for}\ r > R} \\
{1 + r_{S}/R \equiv \psi_{0}\ \text{for}\ r < R} \\
\end{array} \right.
\tag{A4}
\end{equation}

The constant values $V_{0} $ and $\psi_{0} $ in the interior of the shell ensure both flat Minkowski spacetime as represented in conveniently scaled coordinates, as well as continuity of the metric tensor across the shell. The static metric given by Equations (A2) and (A3) will be used as a ``base" metric, i.e., as the zeroth order term of an approximation for a rotating~shell.

If the shell is not static, but slowly rotating with an angular velocity $\omega_{s} $, the line element (A2) must be perturbed to incorporate the effects of the rotation on the geometry. Thus, we consider a perturbation of the line element (A2),
        \begin{equation}
\label{eq:FD53}
ds^{2} = V^{2}dt^{2} - \psi^{4}{({dr^{2} + \ r^{2}{({d\vartheta^{2} + \sin ^{2}\vartheta\left( {d{\phi} - \text{$\Omega$}\left( r \right)dt} \right)^{2}})}})}\ ,
\tag{A5}
\end{equation}
        where $\text{$\Omega$}\left( r \right) $ describes the rotation of the inertial frames. In order to calculate the components of the Einstein tensor from this line-element, it is useful to introduce an orthonormal basis with the following basis~forms,
        \begin{equation}
\label{eq:FD54}
{\omega^{0} = Vdt\ },
\tag{A6}
\end{equation}
\begin{equation}
\label{eq:FD55}
{\omega^{1} = \psi^{2}dr},
\tag{A7}
\end{equation}
\begin{equation}
\label{eq:FD56}
\omega^{2} = r\psi^{2}d\vartheta,\
\tag{A8}
\end{equation}
\begin{equation}
\label{eq:FD57}
{\omega^{3} = r\psi^{2}\sin \vartheta\left( {d{\phi} - \text{$\Omega$}dt} \right)\ }.
\tag{A9}
\end{equation}

Using the Cartan formalism~\cite{B40}, we then find the following contravariant components of the Einstein~tensor,
        \begin{equation}
\label{eq:FD58}
G^{00} = - \frac{4\left( {2\psi^{\prime} + r\psi^{_{''}}} \right)}{r\psi^{5}}\ ,
\tag{A10}
\end{equation}
\begin{equation}
\label{eq:FD59}
G^{11} = \frac{2\left( {2\psi V\psi^{\prime} + 2rV\psi\prime^{2} + \psi^{2}V^{\prime} + 2r\psi\psi\prime V\prime} \right)}{r\psi^{6}V}\ ,
\tag{A11}
\end{equation}
\begin{equation}
\label{eq:FD60}
G^{22} = G^{33} = \frac{2\psi V\psi^{\prime} - 2rV{\psi^{\prime}}^{2} + \psi^{2}V^{\prime} + 2r\psi V\psi^{_{''}} + r\psi^{2}V''}{r\psi^{6}V}\ ,
\tag{A12}
\end{equation}
\begin{equation}
\label{eq:FD61}
G^{03} = G^{30} = \frac{\sin \vartheta}{2\psi^{3}V^{2}}\left( {- 4\psi V\text{$\Omega$}^{\prime} - 6rV\text{$\Omega$}^{\prime}\psi^{\prime} + r\psi\text{$\Omega$}^{\prime}V^{\prime} - r\psi V\text{$\Omega$}''} \right)\ ,
\tag{A13}
\end{equation}
        where
        \begin{equation}
\label{eq:FD62}
\psi^{\prime} = - \frac{r_{S}}{r^{2}}\theta\left( {r - R} \right)\ ,
\tag{A14}
\end{equation}
\begin{equation}
\label{eq:FD63}
\psi^{_{''}} = \frac{2r_{S}}{r^{3}}\theta\left( {r - R} \right) - \frac{r_{S}}{r^{2}}\delta\left( {r - R} \right)\ ,
\tag{A15}
\end{equation}
\begin{equation}
\label{eq:FD64}
V^{\prime} = \frac{2r_{S}}{\left( {r + \alpha} \right)^{2}}\theta\left( {r - R} \right)\ ,
\tag{A16}
\end{equation}
\begin{equation}
\label{eq:FD65}
V^{_{''}} = - \frac{4r_{S}}{\left( {r + r_{S}} \right)^{3}}\theta\left( {r - R} \right) + \frac{2r_{S}}{\left( {r + r_{S}} \right)^{2}}\delta\left( {r - R} \right)\ .
\tag{A17}
\end{equation}

Here, $\theta\left( r \right) $ is the step function, and $\delta\left( r \right) $ is the Dirac delta function. The diagonal components of the Einstein tensor then reduce to the~expressions
        \begin{equation}
\label{eq:FD66}
G^{00} = \frac{4r_{S}}{r^{2}\psi^{5}}\delta\left( {r - R} \right)\ ,
\tag{A18}
\end{equation}
\begin{equation}
\label{eq:FD67}
G^{11} = 0\ ,
\tag{A19}
\end{equation}
\begin{equation}
\label{eq:FD68}
G^{22} = G^{33} = \frac{r_{S}}{2r\psi V}G^{00}\ .
\tag{A20}
\end{equation}

These components are identical to those obtained for a \emph{nonrotating} shell, i.e., the zeroth-order result for which $\omega_{s} = 0 $. In this particular case, Einstein’s field equations then~yield
  \vspace{12pt}
      \begin{equation}
\label{eq:FD69}
{\rho} = T^{00} = \frac{G_{00}}{8\pi} = \frac{r_{S}r^{2}}{2\pi\left( {r + r_{S}} \right)^{5}}\delta\left( {r - R} \right)\ ,
\tag{A21}
\end{equation}
\begin{equation}
\label{eq:FD70}
T^{33} = T^{22} = \frac{G_{22}}{8\pi} = \frac{r_{S}}{2\left( {r - r_{S}} \right)}{\rho}\ ,
\tag{A22}
\end{equation}
        where ${\rho} $ is the mass density of the shell in the rest frame of an element of the shell and $T^{\mu\nu} $ are the components of the stress-energy tensor.

In order to determine the nondiagonal component $G^{03} $ for the rotating shell, we note that Equation (A13) may be written~as
        \begin{equation}
\label{eq:FD71}
G^{03} = - \frac{\sin \vartheta}{2r^{3}\psi^{8}}\left( \frac{r^{4}\psi^{6}\text{$\Omega$}^{\prime}}{V} \right)^{\prime}\ .
\tag{A23}
\end{equation}

In the interior and exterior of the shell, we have $T^{\mu\nu} = 0 $. Accordingly, the corresponding field equation~becomes
        \begin{equation}
\label{eq:FD72}
\left( \frac{r^{4}\psi^{6}\text{$\Omega$}^{\prime}}{V} \right)^{\prime} = 0\ r \neq R\ ,
\tag{A24}
\end{equation}
        which may be integrated to give
        \begin{equation}
\label{eq:FD73}
\text{$\Omega$}_{\pm}^{\prime} = \frac{K_{\pm}V\psi^{2}}{\left( {r\psi^{2}} \right)^{4}}\ ,
\tag{A25}
\end{equation}
        where $K_{-} $ and $K_{+} $ denote constants of integration in the interior and exterior, respectively. In the interior region $r < R $, we have $\psi = \psi_{0} $ and $V = V_{0} $, yielding
        \begin{equation}
\label{eq:FD74}
\text{$\Omega$}_{-} = - \frac{K_{-}V_{0}}{3r^{3}\text{$\psi$}_{0}^{6}} + \text{$\Omega$}_{B}\ .
\tag{A26}
\end{equation}

Here, $\text{$\Omega$}_{B} $ is a new integration constant. Since spacetime is empty and flat in this region, the solution must be regular at $r = 0 $, giving $K_{-} = 0 $. In the exterior region $r > R $, integration of Equation (A25)~gives
        \begin{equation}
\label{eq:FD75}
\text{$\Omega$}_{+}\left( r \right) - \text{$\Omega$}_{+}\left( {r = \infty} \right) = - \frac{K_{+}}{3r^{3}\psi^{6}}\ .
\tag{A27}
\end{equation}

Without loss of generality, we may conveniently choose $\text{$\Omega$}_{+}\left( {r = \infty} \right) = 0 $. Then, $\text{$\Omega$}_{+}\left( r \right) $ describes by definition the local rotation of inertial frames with respect to the nonrotating inertial frames at~infinity.

Requiring continuity across the shell, we must now have $\text{$\Omega$}_{-}\left( R \right) = \text{$\Omega$}_{+}\left( R \right) $ from which we deduce that $K_{+} = - 3\left( {R\,\psi_{0}^{2}} \right)^{3}\text{$\Omega$}_{B} $. As a result, we~obtain
        \begin{equation}
\label{eq:FD76}
\text{$\Omega$}\left( r \right) = \left\{ {\begin{array}{c}
{\left( \frac{R\psi_{0}^{2}}{r\psi^{2}} \right)^{3}\text{$\Omega$}_{B}\ \text{for}\ r > R} \\
{\ \text{$\Omega$}_{B}\ \text{for}\ r < R} \\
\end{array}\ .} \right.
\tag{A28}
\end{equation}

The constant $\text{$\Omega$}_{B} $ will be determined later from the field equations in the region containing the shell. Differentiating, we then find~that
        \begin{equation}
\label{eq:FD77}
\text{$\Omega$}^{\prime} = - \frac{3\text{$\Omega$}_{B}\left( R\psi_{0}^{2} \right)^{3}V}{\left( {r^{2}\psi^{3}} \right)^{2}}\theta\left( {r - R} \right)\ ,
\tag{A29}
\end{equation}
        and
        \begin{equation}
\label{eq:FD78}
\text{$\Omega$}^{_{''}} = 3\text{$\Omega$}_{B}\left( R\psi_{0}^{2} \right)^{3}\left\lbrack {\frac{2\left( {2r^{2} - 4r_{S}r + r_{S}{}^{2}} \right)}{r^{7}\psi^{8}}\theta\left( {r - R} \right) - \frac{V}{\left( {r^{2}\psi^{3}} \right)^{2}}\delta\left( {r - R} \right)} \right\rbrack\ .
\tag{A30}
\end{equation}

Finally, upon substituting Equations (A29) and (A30) in Equation (A13), the expression of the nondiagonal component of the Einstein tensor~becomes
        \begin{equation}
\label{eq:FD79}
G^{03} = \frac{3\text{$\Omega$}_{B}\left( R\psi_{0}^{2} \right)^{3}\sin \upsilon}{2r^{3}\psi^{8}}\delta\left( {r - R} \right)\ .
\tag{A31}
\end{equation}

In order to calculate the rotation $\text{$\Omega$}\left( r \right) $ of inertial frames induced by the rotating shell, we must now consider the stress-energy tensor for the shell. Following Brill and Cohen~\cite{B14}, we assume that the stress-energy tensor has the~form
        \begin{equation}
\label{eq:FD80}
T^{\mu\nu} = {\rho} u^{\mu}u^{\nu} + \sum\limits_{i,j = 1}^{3}t^{ij}v_{(i)}^{\mu}v_{j}^{\nu}\ .
\tag{A32}
\end{equation}

Here, ${\rho} $ is the mass density in the rest frame of the shell, $u^{\mu} $ the four-velocity of an element of the shell and $v_{(i)}^{\mu} $ form a triad of orthonormal vectors spanning the hypersurface orthogonal to $u^{\mu} $. $T^{\mu\nu} $~then satisfies the requirement that the momentum density $T^{i0} $ vanishes in the rest frame of the matter. The coordinate basis components of the four-velocity of an element of the shell rotating with angular velocity $\frac{d{\varphi}}{dt} = \omega_{s} $ may now be calculated from the line element (A2) with $dr = d\vartheta = 0 $,~giving
        \begin{equation}
\label{eq:FD81}
{\widetilde{u}}^{0} = \frac{dt}{d\tau} = V\left( {1 - \sigma^{2}} \right)^{- 1/2}\ ,
\tag{A33}
\end{equation}
\begin{equation}
\label{eq:FD82}
{\widetilde{u}}^{1} = {\widetilde{u}}^{2} = 0\ ,
\tag{34}
\end{equation}
\begin{equation}
\label{eq:FD83}
{\widetilde{u}}^{3} = \omega_{s}{\widetilde{u}}^{0}\ ,
\tag{A35}
\end{equation}
        where
        \begin{equation}
\label{eq:FD84}
\sigma = \frac{r\psi^{2}\sin \vartheta\left( {\omega_{s} - \text{$\Omega$}} \right)}{V}\ .
\tag{A36}
\end{equation}

The orthonormal 1-form basis is related to the coordinate basis by the~transformation
        \begin{equation}
\label{eq:FD85}
\omega^{\nu} = \alpha_{\mu}^{\ \nu}\left( x \right)\ dx^{\mu}\ ,
\tag{A37}
\end{equation}
        where the explicit forms of the $\alpha_{\mu}^{\ \nu}\left( x \right) $ are given in Equations (A6)--(A9). The corresponding dual bases are thus related by the transformation
        \begin{equation}
\label{eq:FD86}
e_{\mu} = \alpha_{\mu}^{\hat{\nu}}\left( x \right)e_{\hat{\nu}}
\tag{A38}
\end{equation}

Here, $e_{\mu} $ and $e_{\hat{\mu}} $ denote the coordinate and orthonormal basis vectors, respectively. The four-velocity may now be expressed in two~ways,
        \begin{equation}
\label{eq:FD87}
\mathbf{u} = {\widetilde{u}}^{\mu}e_{\mu} = {\widetilde{u}}^{\mu}\alpha_{\mu}^{\hat{\nu}}\left( x \right)e_{\hat{\nu}}
\tag{A39}
\end{equation}
        from which it follows that the orthonormal basis components are given by
        \begin{equation}
\label{eq:FD88}
u^{\nu} = \alpha_{\mu}^{\ \nu}\left( x \right){\widetilde{u}}^{\mu}\ .
\tag{A40}
\end{equation}

This yields the~result
        \begin{equation}
\label{eq:FD89}
u^{0} = \left( {1 - \sigma^{2}} \right)^{- 1/2}\ ,
\tag{A41}
\end{equation}
\begin{equation}
\label{eq:FD90}
u^{1} = u^{2} = 0\ ,
\tag{A42}
\end{equation}
\begin{equation}
\label{eq:FD91}
u^{3} = \sigma\left( {1 - \sigma^{2}} \right)^{- 1/2}\ .
\tag{A43}
\end{equation}

The triad of orthonormal vectors are now chosen~as
        \begin{equation}
\label{eq:FD92}
v_{(1)}^{\mu} = \left( {0,\ 1,\ 0,\ 0} \right)\ ,
\tag{A44}
\end{equation}
\begin{equation}
\label{eq:FD93}
v_{(2)}^{\mu} = \left( {0,\ 0,\ 1,\ 0} \right)\ ,
\tag{A45}
\end{equation}
\begin{equation}
\label{eq:FD94}
v_{(3)}^{\mu} = \left( {\sigma,\ 0,\ 0,\ 1} \right)\left( {1 - \sigma^{2}} \right)^{- 1/2}\ .
\tag{A46}
\end{equation}

Inserting these vectors into the general expression (A32) for the stress-energy tensor, where $t^{ij} $ is diagonal, gives the components of the stress-energy tensor of the shell correct to first order in $\omega_{s} $,
        \begin{equation}
\label{eq:FD95}
T^{00} = \frac{{\rho} + \sigma^{2}}{1 - \sigma^{2}} \approx {\rho}\ ,
\tag{A47}
\end{equation}
\begin{equation}
\label{eq:FD96}
T^{22} = \ t^{22},
\tag{A48}
\end{equation}
\begin{equation}
\label{eq:FD97}
T^{33} = \ t^{33},
\tag{A49}
\end{equation}
\begin{equation}
\label{eq:FD98}
T^{03} = \ \left( {{\rho} + t^{33}} \right)\sigma.
\tag{A50}
\end{equation}

Using the zeroth-order result from Equation (A22) for the nonrotating shell, we further~have
        \begin{equation}
\label{eq:FD99}
t^{22} = t^{33} = \frac{{\rho} r_{S}}{2\left( {R - r_{S}} \right)} \equiv {\rho}\beta\ .
\tag{A51}
\end{equation}

Hence, $T^{03} = {\rho}\left( {1 + \beta} \right)\sigma $, where ${\rho} $ is given by the zeroth order result in Equation (A23). Integrating Einstein’s field equation, $G^{03} = 8\pi T^{03} $, across the shell then~yields
        \begin{equation}
\label{eq:FD100}
\int\frac{3\text{$\Omega$}_{B}\left( {R\psi_{0}^{2}} \right)^{3}\sin \vartheta r^{5}}{2\left( {r + r_{S}} \right)^{8}}\delta\left( {r - R} \right)\ dr = \int\frac{4r_{S}\sin \vartheta\left( {1 + \beta} \right)r^{2}\left( {\omega_{s} - \text{$\Omega$}\left( r \right)} \right)}{\left( {r + r_{S}} \right)^{2}\left( {r - r_{S}} \right)}\delta\left( {r - R} \right)\ dr\ ,
\tag{A52}
\end{equation}
        or
        \begin{equation}
\label{eq:FD101}
\frac{3\text{$\Omega$}_{B}R^{2}\sin \vartheta}{2\left( {R + r_{S}} \right)^{2}} = \frac{4r_{S}\sin \vartheta\left( {1 + \beta} \right)R^{2}\left( {\omega_{s} - \text{$\Omega$}_{B}} \right)}{\left( {R + r_{S}} \right)^{2}\left( {R - r_{S}} \right)}\ ,
\tag{A53}
\end{equation}
        where $\text{$\Omega$}\left( R \right) = \text{$\Omega$}_{B} $. Solving this equation for $\text{$\Omega$}_{B} $, we obtain
        \begin{equation}
\label{eq:FD102}
\text{$\Omega$}_{B} = \dfrac{\omega_{s}}{1 + \dfrac{3\left( {R - r_{S}} \right)}{8r_{S}\left( {1 + \beta} \right)}}\ .
\tag{A54}
\end{equation}

Substituting this expression in Equation (A28), we finally arrive at the result of Brill and Cohen~\cite{B14} for the induced rotation rate $\text{$\Omega$}\left( r \right) $ of the inertial frames as determined by the rotating mass of the~shell,
        \begin{equation}
\label{eq:FD103}
\text{$\Omega$}\left( r \right) = \left\{ {\begin{array}{l}
{\dfrac{\left( \dfrac{R\psi_{0}^{2}}{r\psi^{2}} \right)^{3}\omega_{s}}{1 + \dfrac{3\left( {R - r_{S}} \right)}{8r_{S}\left( {1 + \beta} \right)}}\ \text{for}\ r > R} \\
{\dfrac{\omega_{s}}{1 + \dfrac{3\left( {R - r_{S}} \right)}{8r_{S}\left( {1 + \beta} \right)}}\ \text{for}\ r < R} \\
\end{array}\ .} \right.
\tag{A55}
\end{equation}

We are here particularly interested in the dragging angular velocity inside the shell. Inserting the expression (A51) for \emph{$\beta$}, it can be written~as
        \begin{equation}
\label{eq:FD104}
\text{$\Omega$} = \frac{4r_{S}\left( {2R - r_{S}} \right)}{\left( {3R - r_{S}} \right)\left( {R + r_{S}} \right)}\omega_{S}
\tag{A56}
\end{equation}

This is equivalent to the dragging coefficient given in Equation (1).

\bibliographystyle{mdpi}
\renewcommand\bibname{References}

\bibliographystyle{mdpi}

\end{document}